\documentclass[pra,twocolumn,showpacs,preprintnumbers,amsmath,amssymb,floatfix,bibnotes,footinbib]{revtex4}

\usepackage{graphicx}% Include figure files 
\usepackage{dcolumn}%Align table columns on decimal point  
\usepackage{bm}% bold math 
\usepackage{amsmath} 
\def\cm{\mbox{cm$^{-1}$}}
\def\rb{$^{85}$Rb}

\def\E{E_{\textrm{pulse}}}

\begin{document}
\title{
  Dynamical interferences 
  to probe short-pulse 
  photoassociation of Rb atoms\\
  and stabilization of Rb$_2$ dimers
  %Controlling Photoassociation and Stabilization of
  %Rubidium Dimers with Pulsed Lasers
}

\author{Jordi Mur-Petit}%%J 
%\email{jordi.mur@lac.u-psud.fr}
\author{Eliane Luc-Koenig}
\author{Fran\c{c}oise Masnou-Seeuws}
\affiliation{
  Laboratoire Aim\'e Cotton,
  CNRS 
  and Universit\'e Paris-Sud, 
  B\^at.\ 505, % Campus d'Orsay,
  F-91405 Orsay, France }%
%\thanks{
%  %Laboratoire Aim\'e Cotton is 
%  UPR 3321 of CNRS associ\'ee \`a l'Univ.\ Paris-Sud,
%  member of %the
%  F\'ed\'eration Lumi\`ere Mati\`ere (FR 2764)
%  and of %the 
%  Institut Francilien de Recherche sur les Atomes Froids (IFRAF).
%}

\date{\today}


\begin{abstract}
%This theoretical paper analyzes  
We analyze 
the formation of Rb$_2$ molecules with  
short photoassociation pulses applied to a cold $^{85}$Rb sample.
%The  laser field
A pump laser pulse
couples a continuum level of the ground electronic 
state X$^1\Sigma_{\textrm{g}}^+$ with  bound levels in the 
$0_{\textrm{u}}^+$(5S+5P$_{1/2}$) and 
$0_{\textrm{u}}^+$(5S+5P$_{3/2}$) vibrational series.
The nonadiabatic coupling between the two excited channels induces 
time-dependent beatings in the  populations. 
We propose to 
take advantage of 
these oscillations %either 
to design further laser pulses that %to 
probe the photoassociation process via 
%a photo-ionization pulse,
photoionization 
or that optimize the stabilization %step 
in deep levels of the ground state.
\end{abstract}

\maketitle

%%%%%%%%%%%%%%%%%%%%%%%%%%%%%%%%
%%%%%%%%%%%%%%%%%%%%%%%%%%%%%%%%

Making ultracold molecules in the 
lowest vibrational level
$v=0$  of the ground electronic 
state and creating stable molecular condensates is presently 
an important challenge since it opens the road toward
ultracold chemistry~\cite{Heinzen2000,Bala2001}. %aux1}
Schemes based on photoassociation (PA) of ultracold atoms~\cite{Jones2006} 
with cw lasers, %followed by radiative stabilization, 
have been very successful to form molecules in an excited electronic state. 
The latter have been  stabilized into excited 
vibrational levels of the ground electronic 
state~\cite{Fioretti1998,Gabbanini2000}, but not yet into $v=0$
except for the case of RbCs~\cite{Sage2005}. %,Jones2006}.
The possibility of controlling PA by use of short laser pulses has been discussed in 
theoretical papers~\cite{Vala2000,Luc2004,Koch2006a,Shapiro2007} and very recently 
attempted by two experimental groups~\cite{Brown2006,Salzmann2006}, 
both in the rubidium case. Success in such experiments will  create 
a bridge between the two domains of cold matter %~\cite{Jones2006} 
and coherent control, %~\cite{Rabitz2000}, 
where femtosecond (fs) pulses are used to 
%probe the dynamics of chemical reactions and to control the exit channels
control chemical reactions~\cite{Rabitz2000}.
Unfortunately, up to now, PA experiments with fs laser 
pulses have achieved {\em destruction} of the molecules 
already existing in the trap rather than creation of 
additional molecules~\cite{Brown2006,Salzmann2006}.

Finding ways to avoid this destructive effect
is therefore a crucial step in the development of experiments.
A promising route is PA through the
resonant coupling mechanism as realized, with cw lasers, 
in Cs$_2$~\cite{Dion2001} and RbCs~\cite{Kerman2004b}. 
For the case of Rb$_2$, it populates the $0_{\textrm{u}}^+$(5S+5P$_{1/2}$) 
and $0_{\textrm{u}}^+$(5S+5P$_{3/2}$) coupled series. 
This is a textbook example of  
global mixing of two molecular vibrational series
due to 
spin-orbit (SO) coupling %, 
%in the intermediate coupling regime, between two molecular vibrational series, 
%yielding 
and manifested by strong perturbations 
in the Rb$_2$~$0_{\textrm{u}}^+$ fluorescence and cw photoassociation spectra
%as analyzed in Refs.~\cite{Amiot1999,Kokoouline2000,Jelassi2006}
%%%JMP (cf.\ Refs.~\cite{Amiot1999,Kokoouline2000,Jelassi2006}).
(cf.\ Refs.~\cite{spectra,Kokoouline2000}).
%
% The aim of the present paper is to draw attention on the dynamical interferences 
% generated in the coupled $\zerou$(5S+5P$_{1/2,3/2}$)  channels in 
% short-pulse photoassociation experiments, and to show that they 
% provide efficient schemes for probe 
% {\bf and} stabilization~\cite{Koch2006b}. % or storage. 
The aim of the present paper is to draw attention to the 
{\em coherent} character of the time evolution in the coupled excited states
{\em after} the PA pulse, and to the possibility or taking advantage of the subsequent
dynamical interferences to optimize stabilization and avoid the destruction
of the molecules.
We consider a simple model with pulses in the picosecond range 
which populate a limited number of bound vibrational levels, 
 and we analyze characteristic time-dependent oscillations that should 
appear in experimental signals.

Time-dependent fluorescence signals manifesting the 
coupling between deeply bound levels of the two series have been 
previously observed by pump-probe spectroscopy~\cite{Zhang2003}.
In this experiment, a molecular beam of Rb$_2$ was excited by 
a 120-fs laser pulse from the $v=0$ level of X$^1\Sigma_{\textrm{g}}^+\equiv$X 
to low ($\sim 5600$~\cm\ below the 5S+5P asymptote)
vibrational levels of the A$^1\Sigma_{\rm u}^+\equiv$A state,
with a classical inner turning point $\sim 8a_0$. 
A delayed pulse, operating %%J in a suitable detection window 
around the outer turning point of the b$^3\Pi_{\textrm{u}}\equiv$b potential (12$a_0$), 
probes the population transferred to this state. 
It presents an oscillatory 
behavior with several characteristic periods of the order of 1~ps.
This is explained  in terms of interferences 
between several paths for the motion of the wavepacket in 
the  excited state due to  the crossing, at
$R_{\rm short} \approx 9.3a_0$, of the potential energy curves 
(hereafter referred to as $V_A$ and $V_b$), 
corresponding to the two Hund's case $a$ A and b states.

\begin{figure}[bth]
  \centering
  \includegraphics[width=0.8\columnwidth,clip=true]{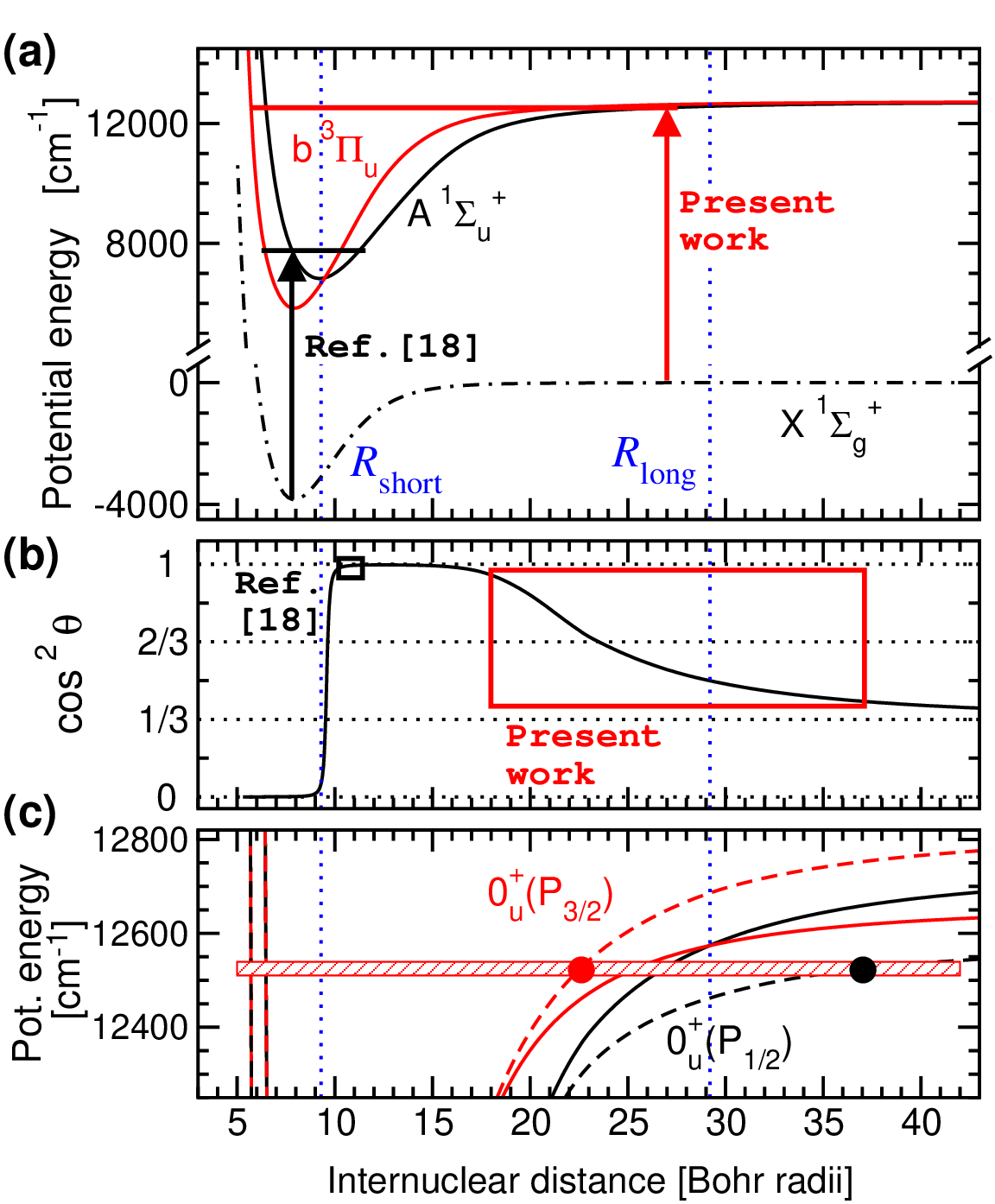}
  \caption{(color online)
    (a) Potential energy curves for the
    ground state X$^1\Sigma_{\textrm{g}}^+$ (dash-dotted line),
    $V_A$ (black, solid) and $V_b$ [red (gray), solid].
%    the diagonal element of the SO coupling has been subtracted from \b.
    (b) $|\textrm{A}\rangle$ component in 
    $|0_{\textrm{u}}^+(\mbox{P}_{1/2})\rangle$ as given by 
    Eqs.~(\ref{eq:angle}).
    (c) Zoom of the energy range where PA takes place,
    showing both diabatic $\{ \mbox{A},\bar{\rm b} \}$ (solid lines) 
    and adiabatic (dashed) potentials.
    The shaded box indicates the PA window in the energy domain
    for the pulse studied.
    The dotted lines are at $R_{\rm short}$ and $R_{\rm long}$,
    and symbols stand for the classical turning points 
    of the levels contributing to the 
    decomposition of the excited wavepacket
    in the uncoupled $0_{\textrm{u}}^+$ basis.
    %excited by the PA pulse.
    %See the text for the meaning of the symbols.
  }
  \label{fig:potentials}
\end{figure}

This crossing can be seen in Fig.~\ref{fig:potentials}(a), 
where we plot $V_A$ and $V_b$ as determined 
%by Bergeman {\em et al.}~\cite{Bergeman2006}.
in Ref.~\cite{Bergeman2006}.
We also show the 
ground-state 
%\gs\
potential obtained by 
matching {\em ab initio} calculations~\cite{Aymar2006} to the 
long-range dispersion potential
$-\sum_n C_n/R^n$~\cite{Marte2002}, and 
adjusting  the  repulsive barrier  to 
reproduce 
%get
the large value of the scattering length~\cite{vanKempen2002}.

In this paper, we consider PA %%J photoassociation 
at a detuning
of $\approx~60~\cm$ below the D$_1$ line ($1~\cm\approx30$~GHz), 
exciting loosely bound vibrational levels of the 
coupled system in the vicinity of the up-to-now unexplored crossing 
at long range $R_{\rm long}\approx 29.2a_0$ [Fig.~\ref{fig:potentials}(c)]. %%J 
The $V_A$ curve crosses 
the ${\bar V_b}$ potential curve 
shifted down from $V_b$ by $\Delta E_{\rm fs}/3=79~\cm$ by %, due to 
the inclusion of the diagonal part of the spin-orbit effective  
Hamiltonian, $H_{SO}$~\cite{Bergeman2006}.
% in the energy of the state $|{\rm b}\rangle$,
%hereafter referred to as diabatic.

%
%
%%J Indeed, inclusion of the spin-orbit effective 
%%J Hamiltonian $H_{SO}$ lowers the energy of the b state by 
%%J a quantity asymptotically $\sim$ 79 \cm, giving rise to
%%J the diabatic curve ${\bar V}_b$, 
%%J that crosses $V_A$ at $R_{\rm long}$.
The diagonalization of $H_{SO}$ within the 
$\{ {\rm A},\bar{ {\rm b} } \}$ subspace, hereafter referred to as {\em diabatic}, 
renders two Hund's case~$c$ $0_{\textrm{u}}^+(\mbox{P}_{1/2})$ and 
$0_{\textrm{u}}^+(\mbox{P}_{3/2})$
{\em adiabatic} curves, correlated with 
the 5S+5P$_{1/2}$ and 5S+5P$_{3/2}$ asymptotes, split by 
$\Delta E_{\rm fs}$. %=237~\cm$.
The mixing angle $\theta (R)$ defines the transformation from the 
diabatic to the adiabatic basis at a given internuclear distance $R$:
\begin{subequations}
\begin{align}
    |0_{\textrm{u}}^+(\mbox{P}_{1/2}) \rangle 
    &= \cos\theta(R) \,\mbox{$|\textrm{A}\rangle$}
      + \sin\theta(R) \;|\bar{\rm b}\rangle\:, \\
    |0_{\textrm{u}}^+(\mbox{P}_{3/2}) \rangle 
    &= -\sin\theta(R)\,\mbox{$|\textrm{A}\rangle$}  
      + \cos\theta(R) \;|\bar{\rm b}\rangle\:.
  \end{align}
  \label{eq:angle}
\end{subequations}
In the adiabatic representation, the two 
excited channels are coupled by radial coupling, governed 
by the derivative $d\theta/dR$.
We display in Fig.~\ref{fig:potentials}(b) the 
$R$-dependence of $\cos^2\theta$.
%, giving the weight of the singlet component 
%in the  $\zerou$(P$_{1/2}$) state. 
There is a sharp variation, corresponding to a singlet-triplet 
change of character, in the region of $R_{\rm short}$, where 
the two diabatic curves cross abruptly.
In the range of distances from 10$a_0$ to 15$a_0$, 
the splitting between the diabatic curves $V_A$ and ${\bar V_b}$ 
is very large compared to $\Delta E_{\rm fs}$, 
%the mixing angle becomes 0, so that 
and $0_{\textrm{u}}^+$(P$_{1/2}$)  has a pure 
singlet character. For $R> 15 a_0$, $\cos^2\theta (R)$ decreases toward
its asymptotic value 1/3, which, due to the 
very similar slopes of $V_A$ and  ${\bar V_b}$ %%J $\bar{V}_b$ 
%in the asymptotic region
at large $R$ [cf. Fig.~\ref{fig:potentials}(c)],
is reached only at distances far beyond  40$a_0$. 
%%J A complete mixing with $\theta = 45^{\circ}$ occurs in the
%%J vicinity of $R_{\rm long}$.
In contrast with the
crossing at $R_{\rm short}$ explored in Ref.~\cite{Zhang2003},
this long-range crossing is not localized.

%%%%%%%%%%%%%%%%%%%%%%%%%%%%%%%%
%%%%%%%%%%%%%%%%%%%%%%%%%%%%%%%%

%{\em The system --}
We perform calculations for a system of \rb\ atoms at a temperature of 
100~$\mu$K as in the usual photoassociation
experiments~\cite{Brown2006,Salzmann2006}. 
The \rb\ isotope is chosen since the manifestation of the resonant coupling
between the two series is more remarkable~\cite{Kokoouline2000}. 
Only $s$-wave scattering and $J=0$ rotational levels are considered 
%%J New:
for simplicity, even though a full treatment of the rotational structure 
will be required to compare with experimental data when they
become available.
%%J
%The three components of the radial wavefunction correspond to the 
%electronic ground state X, %%J (\gs$\equiv$X), 
%and singlet  A %%J(\sing$\equiv$A)
%and triplet b
%excited states. %%J (\trip$\equiv$b).
We consider a %Gaussian 
chirped pulse of duration $\tau_C$, 
that delivers an energy $\E$ uniformly over an area $\sigma$. 
It is centered at time $t_P$ and has a frequency that varies linearly in time,
$ \omega(t) = \omega_L + \chi\cdot(t-t_P)$,
around the carrier frequency $\omega_L$.
%The detuning with 
%respect to the 5S$\leftrightarrow$5P$_{1/2}$ atomic resonance line 
%at $\omega_{\rm at}$ is $\Delta_L=\hbar(\omega_{\rm at}-\omega_L)$. 
The laser is red detuned from the atomic D$_1$ line at $\omega_{\rm D1}$
by $\Delta_L=\hbar(\omega_{\rm D1}-\omega_L)$;
$\chi$ is the linear chirp rate in the time domain. 
%while the spectral bandwidth is $\delta\omega$.
%=4\ln2 / \tau_L \approx 14.7~\cm / \tau_L[{\rm in~ps}]$, %%J .
%$\tau_L$ being the duration of the transform-limited pulse.
The instantaneous intensity $I(t)$ of the pulse involves a Gaussian envelope
%%J \begin{align}
%%J   \!\!\!I(t) &\!=\! \frac{\E}{\tau_C\sigma}
%%J   %f(t) %\:, \\
%%J   %f(t) 
%%J   %\!=\! {\E}_{\rm max}
%%J   \sqrt{\frac{4\ln2}{\pi}}
%%J   \exp\!\!\left[ -4\ln2
%%J       \left(\frac{t-t_P}{\tau_C}\right)^2 \right]
%%J %      \left(\frac{t-t_{\rm pump}}{\tau_C}\right)^2 \right]\!,
%%J \label{eq:pulse}
%%J \end{align}
with a full width at half maximum equal to $\tau_C$~\cite{Luc2004epjd}.
%($\geq\tau_L$), 
%the chirped pulse duration~\cite{Luc2004epjd}.
%. These parameters are related by
%$
%\chi^2\tau_C^4=(4\ln2)^2[(\tau_C/\tau_L)^2-1]
%$.
For this pulse, 98\% of the energy $\E$ is delivered in the
{\em time window} $[t_P-\tau_C,t_P+\tau_C]$%%J
%%J over the illuminated area $\sigma$
~\cite{Luc2004}, %%J.
%%J During this time, 
the instantaneous laser frequency being then
resonant with 
all the excited levels with a binding energy in the range
$[\Delta_L-\hbar|\chi|\tau_C,\Delta_L+\hbar|\chi|\tau_C]$,
which defines a {\em PA  window in energy}%%J .
%%J For the case of a single excited state, 
%%J this can be translated into a {\em PA window in $R$}
%%J by the reflection principle
~\cite{Luc2004epjd}, cf.\ Fig.~\ref{fig:potentials}(c).

We choose a nonperturbative description  for the dynamics of the  PA process
and of the vibration of the molecules, 
and solve the time-dependent coupled Schr\"odinger equations~\cite{Luc2004,Koch2006a} 
to compute the wavepacket motion in the ground X and the excited A and ${\bar{\rm b}}$ 
states. The X and A channels are coupled by 
%%J channel, with a time dependent coupling %that reads 
$ \hbar\Omega(t) = -\sqrt{ I(t) / (2c\epsilon_0) }
  %\sqrt{\tau_L / \tau_C}
  {\cal D}(R)
%  a(R) {\cal D}
$,
where $c$ and $\varepsilon_0$ are the speed of light and 
permittivity of vacuum, 
%${\cal D}= 5.201ea_0^2$ ($e=$proton charge) is the 
%relevant atomic transition dipole moment 
%and $a(R)$ accounts for the $R$-dependence of the molecular EDM;
%at large distances $a(R)=\sqrt{2/3}$.
and ${\cal D}(R)$ is the molecular transition dipole moment;
at large distances ${\cal D}(R)=4.245ea_0$
($e=$ proton charge, $a_0=$ Bohr radius).
%We apply the rotating wave approximation on the instantaneous frequency
%to eliminate from the dynamical equations the rapidly oscillating 
%terms~\cite{Luc2004,Koch2006a} 
A radial grid up to $30\,000a_0$ has been built 
with the mapped grid method~\cite{Willner2004} 
to faithfully represent the initial state 
as a stationary scattering level in X, 
and the vibrational wavefunctions in the excited states.
%A Chebychev expansion of the evolution operator is used
%for the time propagation as in Ref.~\cite{Luc2004}.
The time propagation was performed as in Ref.~\cite{Luc2004}.

%%%%%%%%%%%%%%%%%%%%%%%%%%%%%%%%
%%%%%%%%%%%%%%%%%%%%%%%%%%%%%%%%

%%J {\em Numerical treatment --}
%%J The low temperature of the experimental systems 
%%J %($T\approx100~\mu$K)
%%J demands the use of a very large simulation box 
%%J for the faithful inclusion as box-states of the continuum states 
%%J above threshold~\cite{Luc2004epjd}.
%%J To this end, a numerical grid up to $L=30\,000a_0$ 
%%J has been built with the Mapped Grid Method~\cite{Willner2004}
%%J and the dynamical equations have been time-propagated 
%%J as in Ref.~\cite{Luc2004}.

%%%%%%%%%%%%%%%%%%%%%%%%%%%%%%%%
%%%%%%%%%%%%%%%%%%%%%%%%%%%%%%%%

From now on, we concentrate on a short PA pulse, red-detuned 
by $\approx60~\cm$ from the  D$_1$ line  
(corresponding to $\lambda_{\rm pump}=798$~nm). 
The bandwidth is set to %$\delta\omega\approx15~\cm$
15~\cm, large enough  to resonantly couple 
the initial state $|{\rm X},E=98.85~\mu{\rm K}\rangle$ with
two levels of the $0_{\textrm{u}}^+(\mbox{P}_{3/2})$ series and 
13 levels of the $0_{\textrm{u}}^+(\mbox{P}_{1/2})$ series, 
but small enough to avoid population of continuum levels in the excited state
[cf.\ Fig.~\ref{fig:potentials}(c)].
The pulse is centered at $t_P=50$~ps, with $\tau_C=10$~ps,
$\chi=4.41\times10^{-2}$~ps$^{-2}$, %%J $\chi=1.47$~\cm~ps$^{-1}$,
and $\E=41$~nJ focused on $\sigma=2.8\times10^{-3}$~cm$^2$.
%The choice of $\chi>0$ enables to maximize the population transfer~\cite{Koch2006b}.
We choose $\chi>0$ to maximize the population transfer~\cite{Koch2006b}.
%%J We set the bandwidth to $\delta\omega\approx15~\cm$
%%J to avoid populating the dissociative levels in the excited states.
%%J The corresponding PA window in energy is indicated 
%%J by the shaded box in Fig.~\ref{fig:potentials}(c).
%For the simulation,
%a numerical grid up to $30\,000~a_0$ has been built with the
%Mapped Grid Method~\cite{Willner2004} to faithfully represent
%the scattering states, and the dynamical equations
%have been propagated as in Ref.~\cite{Luc2004}.

%%%%%%%%%%%%%%%%%%%%%%%%%%%%%%%%
%%%%%%%%%%%%%%%%%%%%%%%%%%%%%%%%

\begin{figure}[tb]
  \centering
  \includegraphics[width=0.85\columnwidth,clip=true]{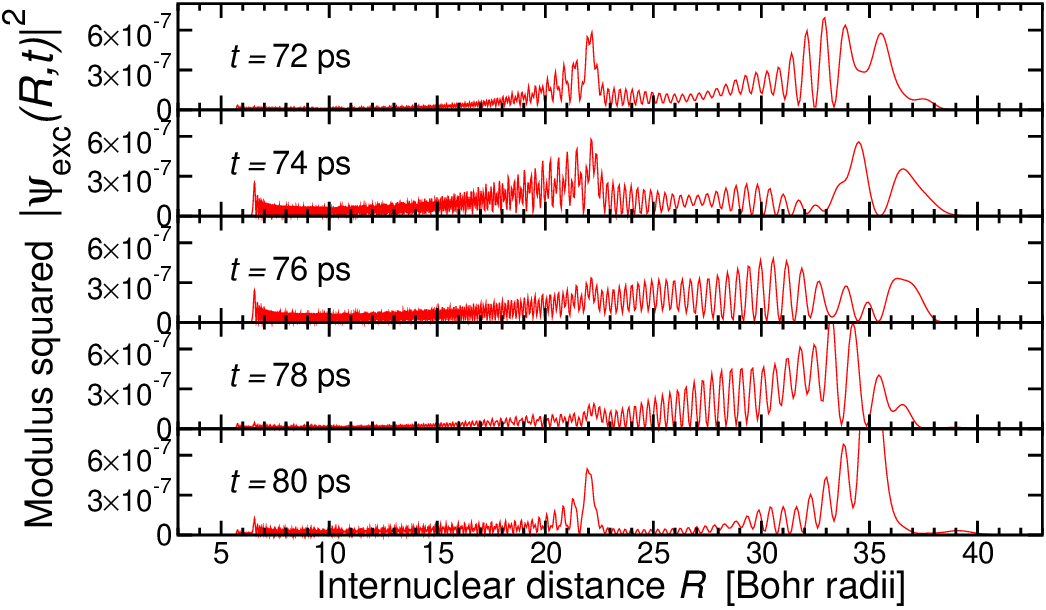}
  \caption{
    (color online) 
    Evolution of the 
    wavepacket in the excited channels
    $\Psi_{\rm exc}(R,t)$
    after the PA pulse has finished.
  }
  \label{fig:evolution}
\end{figure}

%{\em Results --}
We present in Fig.~\ref{fig:evolution} the evolution of the population 
in the excited states, %%J $\Psi_{\rm exc}(R,t)$,
$|\Psi_{\rm exc}(R,t)|^2
=|\Psi_{\rm A}(R,t)|^2+|\Psi_{\bar{{\rm b}}}(R,t)|^2$,
well after the PA pulse has finished.
The PA probability for a single pair of atoms 
is $4.21\times10^{-6}$, and most excited population 
concentrates around two peaks at $22a_0$ and 
$37a_0$. The relative importance of these peaks changes in time. 
For example, the amplitude of the peak at $22a_0$ oscillates with a period $T^*\approx8$~ps.
%%J that is present at $t=72$~ps,
%%J has disappeared at $t=76$~ps, but reappears at $t=80$~ps. 
%%J Thus, a period  
%%J $T^*\approx8$~ps 
%$T_{\rm beat}\approx8$~ps 
%%J can be assigned to these oscillations.

%

This beating results from the interferences between the population 
of the stationary levels $v'$ of the coupled system of excited states. 
Indeed, in the present case, the coupling is such that neither the diabatic 
nor the adiabatic stationary vibrational levels constitute a good basis 
for the analysis. 
Here, the coupled states $|\phi_{v'}^{\rm coup}\rangle$, 
which have components on the two electronic channels,
are labeled by $v'$ according to their increasing energy $E_{v'}$.
In this basis, the decomposition of the two-component wavepacket reads
$\Psi_{\rm exc}(R,t)
=\sum_{v'}c_{v'}\phi_{v'}^{\rm coup}(R) \exp(-iE_{v'}t/\hbar)$,
where each $c_{v'}$ as well as the total excited population 
%in the excited potentials 
are constant after the pulse.

%This behavior can be understood in terms of %%J quantum 
%interferences 
%between the %%J various 
%stationary 
%levels that form the wavepacket.
%In general, t
%%J The excited wavepacket can be decomposed in the
%%J basis formed by the vibrational eigenfunctions 
%The SO coupling between the A and b states makes
%the diabatic basis inappropriate for this analysis. 
%However, the strong non-adiabatic effect indicated by the smooth
%behavior of $\theta(R)$ close to $R_{\rm long}$ render the uncoupled
%adiabatic basis also not good. In fact, it is necessary to work
%with a basis that incorporates such couplings. 

%$\{ \phi_{v'}^{\rm coup} \}$ 
%%J of the {\em coupled} \zerou\ potentials, 

%%J where the summation is over the 
%%J %levels inside the PA window defined above,
%%J %that in the present case comprises $\sim 15$ levels
%%J %of energies $E_{v'}$.
%%J $\sim15$ levels of energies $E_{v'}$ inside the energy
%%J PA window defined above.

\begin{figure}[tb]
  \centering
  \includegraphics[width=0.85\columnwidth,clip=true]{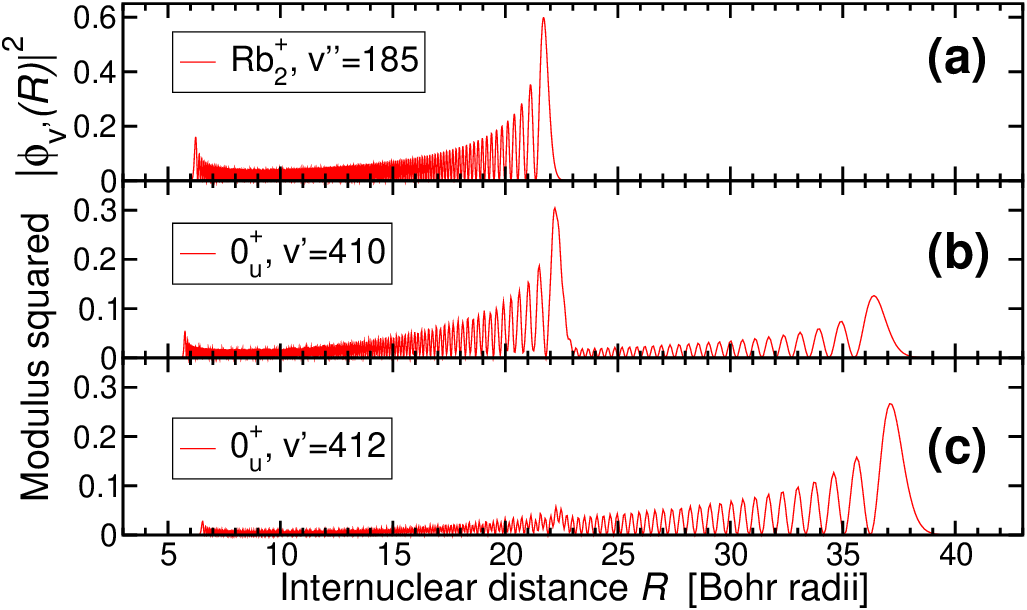}
  \caption{
    (color online)
    Stationary wavefunctions of
    (a) the Rb$_2^+$(X$^2\Sigma_{\rm g}^+$) level bound by 264~\cm; 
    (b) the $0_{\textrm{u}}^+$ level 54~\cm\ below D$_1$;
    and
    (c) the $0_{\textrm{u}}^+$ level 51~\cm\ below D$_1$.
  }
  \label{fig:resonant}
\end{figure}

We show in Figs.~\ref{fig:resonant}(b) and~\ref{fig:resonant}(c)
the radial density $|\phi_{v'}^{\rm coup}(R)|^2$ 
of two stationary states present in the excited wavepacket.
The lower panel corresponds to a level 
%with $\Delta_L=51~\cm$.
$51~\cm$ below the 5S+5P$_{1/2}$ limit. 
It turns out to be ascribable to a ``$0_{\textrm{u}}^+(\mbox{P}_{1/2})$ level'', 
as most of its probability density is accumulated close to the outer turning point 
$R_{\rm ext,1/2}\approx 37a_0$, corresponding to   
this potential and this energy [black circle in Fig.~\ref{fig:potentials}(c)].
%%J More unusual is t
The state in Fig.~\ref{fig:resonant}(b), which corresponds to a level 
%with $\Delta_L=54~\cm$. 
54~\cm\  below D$_1$, %5S+5P$_{1/2}$, %%J. This 
is a typical example of a
``resonant $0_{\textrm{u}}^+(\mbox{P}_{3/2})$ level''~\cite{Kokoouline2000}:
it has a 55\% weight in the adiabatic P$_{3/2}$ channel,  and %it 
features two maxima of probability, corresponding to the turning points 
$R_{\rm ext,1/2}$ and $R_{\rm ext,3/2}\approx 22a_0$ 
[red (gray) circle in Fig.~\ref{fig:potentials}(c)]
in $0_{\textrm{u}}^+(\mbox{P}_{1/2})$ and $0_{\textrm{u}}^+(\mbox{P}_{3/2})$ 
respectively.
As the instantaneous frequency of the chirped pulse is resonant
with $\sim$15 levels in the coupled $0_{\textrm{u}}^+$ series,
$\Psi_{\rm exc}$ %%J the excited wavepacket 
will have components on stationary levels of both types, 
%thus explaining 
which explains the double-peak structure observed in Fig.~\ref{fig:evolution}.

Now, it is easy to show that the  population density 
at an internuclear distance $R$
has a time dependence determined by {\em all} the beating
frequencies $\omega_{ij}=(E_i-E_j)/\hbar$, where $E_i$ and $E_j$ are the 
energies of those levels for which $c_{v'}\neq 0$.
The largest weights in the decomposition correspond
to the levels of the coupled basis labeled by 
$v'=406$ and $408$%%J . These are 
, two ``$0_{\textrm{u}}^+(\mbox{P}_{1/2})$'' levels
separated by $\Delta E\approx4.24~\cm$, which corresponds to a
beating of $T_{\rm beat}\approx7.85$~ps, 
in agreement with $T^*$ 
observed in Fig.~\ref{fig:evolution}.

%%J One could also project the excited wavepacket 
%%J on the uncoupled 
%%J basis formed by the eigenstates
%%J of $V_A$ and $\bar{V}_b$.
%%J In this case, many levels contribute, in particular, 
%%J levels {\em outside} the energy PA window %defined by 
%%J of the pulse.
%%J The largest contributions actually come from A and $\bar{\rm b}$ 
%%J levels %whose 
%%J with classical turning points %are located 
%%J at
%%J $R_{\rm ext,1/2}$ and $R_{\rm ext,3/2}$,
%%J which correspond to the levels indicated by squares %(or circles)
%%J in Fig.~\ref{fig:potentials}(c).
%%J In fact, for the two strongly coupled A and $\bar{\rm b}$ states, 
%%J a PA window cannot be defined in energy nor in $R$.

%%J \begin{figure}[tb]
%%J   \centering
%%J   \includegraphics[width=0.85\columnwidth,clip=true]{resonant-levels}
%%J   \caption{
%%J     (color online)
%%J     Stationary wavefunctions of
%%J     (a) the Rb$_2^+$(X$^2\Sigma_{\rm g}^+$) level bound by 264~\cm; 
%%J     (b) the \zerou\ level 54~\cm\ below D$_1$;
%%J     and
%%J     (c) the \zerou\ level 51~\cm\ below D$_1$.
%%J   }
%%J   \label{fig:resonant}
%%J \end{figure}

\begin{figure}[h!]
  \centering
  \includegraphics[width=0.85\columnwidth,clip=true]{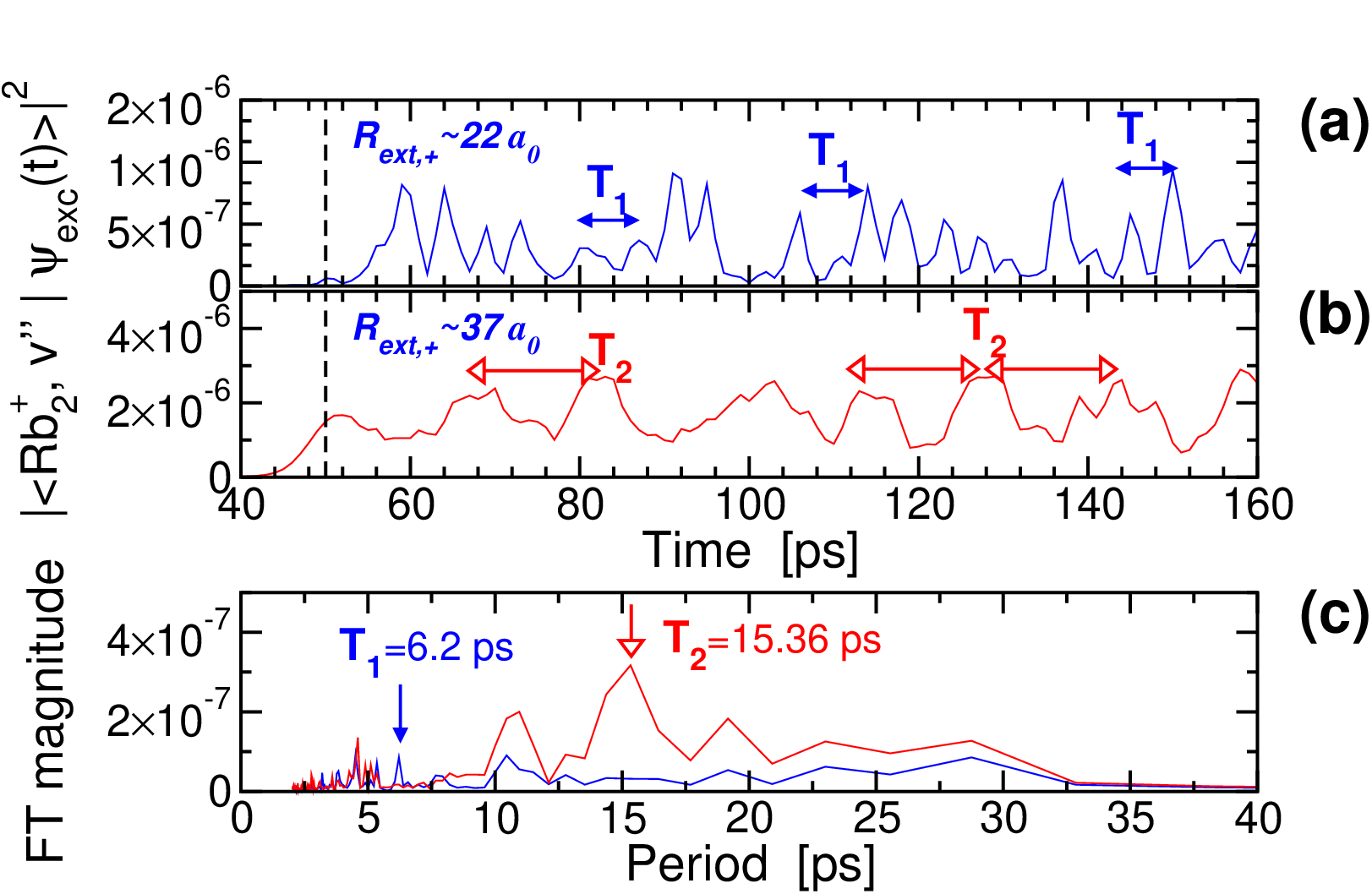}%oscillations}%-ion-stab}
  \caption{
    (color online) 
    Evolution of the 
    overlap of $|\Psi_{\rm exc}(t)\rangle$ with
    (a) 10 Rb$_2^+$(X$^2\Sigma_{\rm g}^+$) levels 
    with $R_{\rm ext,+}\approx 22a_0$;
    (b) 16 Rb$_2^+$(X$^2\Sigma_{\rm g}^+$) levels 
    with $R_{\rm ext,+}\approx 37a_0$.
    (c) Fourier transform of signals in (a) [blue (black) line] 
    and (b) [red (gray) line] in the time window $t\in[70,300]$~ps.
%    (c) 17 Rb$_2$(X$^1\Sigma_{\textrm{g}}^+() levels with $E_{\rm bind}\lesssim 24~\cm$;
%    (d) 28 Rb$_2$(X$^1\Sigma_{\textrm{g}}^+$) levels with $E_{\rm bind}\approx 1300-2500~\cm$.
    The dashed vertical line stands for $t_P=50$~ps
    and arrows indicate the most relevant timescales.
  }
  \label{fig:oscil}
\end{figure}

The observed oscillation in the population density close to 
%a particular $R$ 
$R_{{\rm ext,3/2}}$
can be probed experimentally 
with a suitable laser pulse that ionizes
the molecule and coherently populates several levels of 
the lowest Rb$_2^+({\rm X}^2\Sigma_{\rm g}^+)$
potential~\cite{Aymar2003}
with classical turning point $R_{\rm ext,+}\approx 22a_0$.
The stationary wavefunction of one such level is plotted in
Fig.~\ref{fig:resonant}(a).
Assuming that the ionization process is a vertical transition,
the population transfer to the ionic channel
will concern the part of the wavepacket that is close 
to $R_{\rm ext,+}$, {\em i.e.}, the peak around $R_{\rm ext,3/2}$.
Indeed, we show in Fig.~\ref{fig:oscil}(a)
the time evolution of the overlap
$\sum_{v''} 
  |\langle {\rm Rb}_2^+({\rm X}^2\Sigma_{\rm g}^+), v'' |
  \Psi_{\rm exc}(t) \rangle|^2
$
for the levels $v''\in[185,194]$ that have 
$R_{\rm ext,+}\approx (21-23)a_0$ and can be populated using a 
laser with central wavelength $\lambda_{\rm probe}=479.8$~nm and  
bandwidth $\approx 30~\cm$.
(The overlap is calculated with both A and b components 
of the excited wavepacket, as no selection rules apply to the 
ionizing transition.) % while the emitted electron is disregarded.}
The signal has a rich structure of peaks and troughs. 
In the Fourier transform (FT) of a longer-duration signal 
[cf.\ Fig.~\ref{fig:oscil}(c)], one observes a characteristic time 
$T_1=6.2$~ps (and its multiples), which is identified with the beating period 
between the levels $v'=407$ [a ``$0_{\textrm{u}}^+(\mbox{P}_{1/2})$'' level] 
and $v'=410$ [a ``resonant $0_{\textrm{u}}^+(\mbox{P}_{3/2})$'' level, 
cf.\ Fig.~\ref{fig:resonant}(b)].
Thus, $T_1$ is a characteristic time 
of the P$_{1/2}$-P$_{3/2}$ population transfer.
Quite remarkably, a different ionizing pulse
designed to probe the wavepacket close to 37$a_0$
($\lambda_{\rm probe}=469.3$~nm) would render a signal 
dominated by $T_2=2T_{\rm beat}=15$~ps, 
the beating period between the two ``$0_{\textrm{u}}^+(\mbox{P}_{1/2})$'' 
levels $v'=407,408$ [cf.\ Fig.~\ref{fig:oscil}(b) and~\ref{fig:oscil}(c)]:
as the classical turning point of the $0_{\textrm{u}}^+(\mbox{P}_{3/2})$ 
components is $22a_0$, 
where the Rb$_2^+$ vibrational wavefunctions 
have no appreciable amplitude,
almost no effect of the 
$0_{\textrm{u}}^+(\mbox{P}_{1/2})$-$0_{\textrm{u}}^+(\mbox{P}_{3/2})$ 
coupling is expected to show up at this distance.
%$T_1$ and $T_2$ characterize each ionization signal
%%even if they may not be the highest peaks,
%because they do {\em not} appear in the FT of the other signal:
%the beating between the two 0$_\textrm{u}^+$ states arises from
%their populations {\em and} from constructive interferences %between the different ionization paths.
Of course, some peaks are visible in Fig.~\ref{fig:oscil}(c) that
correspond to other beating periods $\omega_{ij}^{-1}$.
Nevertheless, $T_1$ and $T_2$ can be used as fingerprints of the two
different ionization processes, as each of them appears only in one signal.

\begin{figure}[tb]
  \centering
  \includegraphics[width=0.85\columnwidth,clip=true]{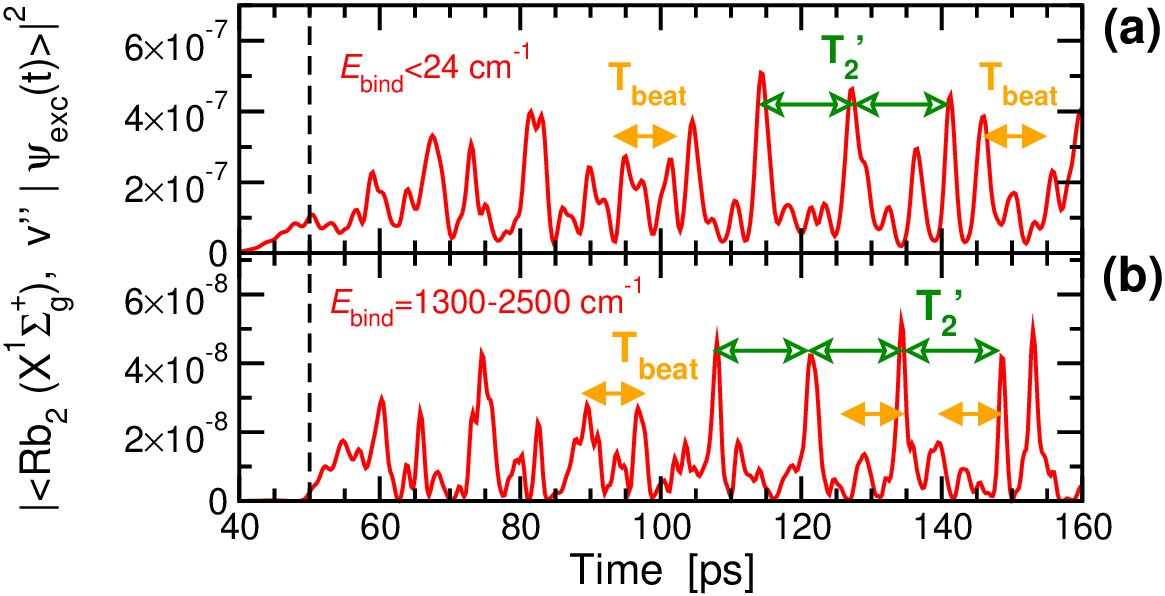}
  \caption{
    (color online)
    Evolution of the 
    overlap of $|\Psi_{\rm exc}(t)\rangle$ with
    (a) 17 Rb$_2$(X$^1\Sigma_{\textrm{g}}^+$) levels with $E_{\rm bind}\lesssim 24~\cm$;
    (b) 28 Rb$_2$(X$^1\Sigma_{\textrm{g}}^+$) levels with $E_{\rm bind}\approx 1300-2500~\cm$.
  }
  \label{fig:NEW}
\end{figure}

We next analyze the dynamics after a dump pulse toward
bound vibrational levels of %%J the electronic ground state, 
Rb$_2$(X$^1\Sigma_{\textrm{g}}^+$). 
Note that due to selection rules, this transition is selecting 
the A$^1\Sigma_{\textrm{u}}^+$ component in the excited wavepacket. 
The results are shown in Fig.~\ref{fig:NEW}(a) for the vibrational 
levels with binding energy $E_{\rm bind}\lesssim 24~\cm$,
and in Fig.~\ref{fig:NEW}(b) for the levels with 
$1300~\cm \leq E_{\rm bind} \leq 2500~\cm$, 
which have a reasonable Franck-Condon factor 
with $\Psi_{\rm exc}$
and are reachable by a fs dump pulse ($\lambda_{\rm dump}=708$~nm).
In both cases, $T_{\rm beat}$ and 
the mean value of the vibrational period in the pure
$0_{\textrm{u}}^+(\mbox{P}_{1/2})$ potential 
{\em in this energy range,}
$T_{2}'\approx 13$~ps, 
have an important role,
and can be used to determine the best timing for the dump pulse.
%%J (The stabilization is more efficient toward the higher-lying levels 
%%J as the corresponding outer turning points are in the 
%%J range $[R_{\rm ext,3/2},R_{\rm ext,1/2}]$.)

%%%%%%%%%%%%%%%%%%%%%%%%%%%%%%%%
%%%%%%%%%%%%%%%%%%%%%%%%%%%%%%%%

In summary, we have studied the dynamics of Rb$_2$~$0_{\textrm{u}}^+$ molecules 
created by a picosecond PA pulse from cold \rb\ atoms.
The excited wavepacket spans $\sim15$ vibrational levels in 
the coupled $0_{\textrm{u}}^+$ basis and presents two maxima of probability 
at $R_{\rm ext,1/2}$ and $R_{\rm ext,3/2}$.
The subsequent dynamics shows quantum interferences, % effects, 
with a beating of the population close to $R_{\rm ext,3/2}$.
%%J We have argued that t
This beating can be monitored by 
photoionizing the wavepacket with a laser with 
well-defined energy spectrum, which in practice
defines a window of internuclear distances 
whose density probability is probed.
The corresponding signal and its Fourier transform 
can serve as identification tool
for ongoing PA experiments~\cite{Salzmann2006,Brown2006}.
%
%We have also analyzed the time-dependence in a pump-dump %sequence to form ground-state molecules in deep bound levels 
%where they may be considered as ``stored''. 
%In an experiment with a high repetition rate of pump-dump pulses,
%a pump pulse would {\em not} dissociate 
%the molecules stabilized by the previous dump pulse, but rather %transfer population to 
%%
%% the {\em lower} levels of the A$^1\Sigma_u^+$ curve, 
%% where the 1~ps beating period~\cite{Zhang2003} 
%% could be exploited to transfer back population to even lower %levels 
%% of the ground state.
%%
%%{\bf
%%%levels bound by $\sim$1650~\cm. 
%still deeply bound vibrational levels.
%From these levels, a set of conveniently delayed two-color %pump-dump pulses could be exploited to transfer back population
% to even lower levels of the
%ground state~(see also~\cite{Shapiro2007}).
%%}
%%
%% The efficiency of such a series of conveniently delayed pulses %should 
%% be compared with the STIRAP mechanism recently %proposed~\cite{Shapiro2007}, 
%% which uses pulses with much larger detunings and smaller %energies.
%%
We have also analyzed a pump-dump pair of pulses to form deeply-bound
ground-state molecules. The resulting time dependence would allow,
in an experiment with a sequence of pump-dump pairs at a high repetition rate,
to choose the delay between consecutive pairs, so that a pump pulse
would not dissociate the molecules created by the preceding dump pulse.
Moreover, these molecules could be further transferred to even deeper levels
by other pairs of pulses (see also Ref.~\cite{Shapiro2007}).
We studied here a pulse with a relatively large detuning 
to avoid coupling to the continuum levels in the excited
channels. 
In the future, for a closer comparison with experiments, 
femtosecond PA pulses with smaller detunings and a mask 
to cut the blue part of the pulse spectrum that transfers 
populations to the continuum, will be analyzed.
Also, we will further study the role of SO coupling
in $^{87}$Rb and $^{133}$Cs samples.

%%%%%%%%%%%%%%%%%%%%%%%%%%%%%%%%
%%%%%%%%%%%%%%%%%%%%%%%%%%%%%%%%

%\acknowledgments
We thank T. Bergeman, M. Aymar, and S. Azizi for providing 
us the potentials in Refs.~\cite{Bergeman2006} 
and~\cite{Aymar2003,Aymar2006}, respectively.
Fruitful discussions with C. Koch, E. Dimova, O. Dulieu, R. Kosloff, 
A. Montmayrant, L. Pruvost, I. Walmsley, and M. Weidem\"uller
are also gratefully acknowledged.
This work was partially supported by the 
EC~Research Training Network ``Cold Molecules''
(contract no.~HPRN-CT-2002-00290).
Laboratoire Aim\'e Cotton is 
UPR 3321 of CNRS, associ\'ee \`a l'Universit\'e Paris-Sud,
member of %the
F\'ed\'eration Lumi\`ere Mati\`ere (FR 2764)
and of %the 
Institut Francilien de Recherche sur les Atomes Froids (IFRAF).

\bibliographystyle{apsrev}
\bibliography{biblio}

\end{document}